\documentclass[useAMS,usenatbib]{mn2e}
\usepackage{graphicx}
\title[Revised Gyro-Age for M\,67]{A Revised Gyro-Age for M\,67 from {\it Kepler/K2-Campaign-5} Light Curves}
\author[G. Gonzalez]{Guillermo Gonzalez$^{1}$\thanks{E-mail:
ggonzalez@bsu.edu}\\
$^{1}$Ball State University, Department of Physics and Astronomy, 2000 W. University Ave., Muncie, 
IN 47306}
\begin{document}

\date{Accepted ??. Received ??; in original form ??}

\pagerange{\pageref{firstpage}--\pageref{lastpage}} \pubyear{??}

\maketitle

\label{firstpage}

\begin{abstract}
We revisit the photometric variability of stars in the M\,67 field using {\it Kepler/K2-Campaign-5} light curves. In our previous work, we limited the search area around M\,67 to that of a recent ground-based study. In the present work, we expand the search area and apply a more rigorous period-finding algorithm to determine the rotation periods of 98 main sequence cluster members from the same data. In addition, we derive periods of 40 stars from the K2SC detrended light curves. We determine the mean period of single sun-like main sequence cluster members to be $29.6 \pm 0.6$ d. Assuming the periods correspond to stellar rotation, the corresponding mean gyro-age is $5.4 \pm 0.2$ Gyr.
\end{abstract}


\section{Introduction}

In \citet{gg16}, hereafter Paper I, we presented analyses of the light curves of 639 stars in the field of M\,67 using data from {\it Kepler/K2-Campaign-5}. We derived a gyro-age of $3.7 \pm 0.3$ Gyr from the rotation periods of 28 sun-like single cluster members. Shortly thereafter, \citet{barnes16}, hereafter B16, published independent analyses of the rotation periods of 20 M\,67 cluster members also using {\it K2} data; they derived a mean gyro-age of $4.2 \pm 0.2$ Gyr.

There are several reasons to conduct another analysis of the rotation periods of stars in M\,67 using {\it K2} data. First, while these two age estimates are consistent with each other, we would like to track down the source of the 0.5 Gyr age difference and try to arrive at an improved estimate. Second, both studies only examined a subset of the M\,67 member stars observed during {\it Kepler/K2-Campaign-5}. B16 limited their full gyro-age analysis only to 20 cluster members, all within 25 arc minutes of the cluster center and outside the inner 10 arc minutes. In Paper I we restricted the sample to the same region of M\,67 observed by \citet{nard16} in their extensive ground-based study. Third, for nearly half the stars in common between Paper I and B16, the derived rotation periods are very different.

Unlike studies conducted on photometry collected during the original {\it Kepler} four year mission, analyses of stellar rotation based on {\it K2} photometry are limited to a mere 80 days or so. This is enough to sample only two to three full rotations for a typical solar age main sequence star. As we showed in Paper I with simulations of rotation period extraction from solar irradiance data, such a short timespan limits the accuracy of the derived period for a sun-like star. Depending on where a particular star is on its activity cycle, it might not even be possible to derive a period for it. Therefore, it becomes necessary to employ as large a sample as possible to arrive at a reliable average period for stars of a given spectral type.

The purpose of the present work is to revisit the photometric variability of M\,67 member stars using the {\it Kepler/K2 Campaign-5} data. Our primary goal is to derive a more accurate mean gyro-age for the cluster. We describe and prepare the data for analysis in Section 2. In Section 3 we discuss the results. We present our conclusions in Section 4.

\section{Data preparation and period analysis}

\subsection{M67 data}

One of our goals in the present study is to improve upon the analysis of Paper I by increasing the number of M\,67 members in our sample. In order to do so, we dropped one of the criteria adopted in that study, namely, that a star in the {\it K2} input catalog be included in the M\,67 field observed by \citet{nard16}. As shown in Figure 1 of Paper 1, their field includes stars up to about half a degree from the cluster center in the E and W directions, but their coverage is only about half as far in the N and S directions. In the following subsections we analyze the K2 light curves from the same database we used in Paper I over this larger region, as well as light curves from other K2 databases.

\subsection{Reanalysis of light curves from Paper I}

In this section restrict our analysis of photometric variability to the same light curves archive we employed in Paper I, using the PDCSAP\_FLUX values.\footnote{https://archive.stsci.edu/k2/} We include all stars from the {\it K2} input catalog within a radius of one degree from the center of M\,67 and with a {\it Kepler} magnitude {\bf brighter than 21.} This yielded 988 stars. As in Paper I, we determined the ``optimal'' period using the python code gatspy\footnote{http://www.astroml.org/gatspy/} for each light curve. We also placed each light curve in one of the subjective categories (see Paper I). This step yielded 441 stars within the ``rot'' category, which yields the most reliable periods according to our analysis of solar data in Section 2.5.

In order to be more successful at selecting the true rotation period for each star, we also employed phase dispersion minimization (PDM).\footnote{We employed a Python implementation of PDM within the PyAstronomy collection at http://www.hs.uni-hamburg.de/DE/Ins/Per/Czesla/PyA/PyA/index.html} PDM arrives at the period which minimizes the variance in a phased time series \citep{stel78}. We describe how we arrived at the final period for each star from the gatspy optimal period and the PDM period in Section 2.5. Applying these selection criteria to the 441 stars in the ``rot'' category, our sample is reduced to 271 stars (27 percent of the original 988 stars). We apply additional culling steps below only to this subset of our sample.

Based on our membership assignments in Table 2 of Paper I for the stars in common with the working sample, we removed another 81 stars as likely nonmembers of M\,67. This leaves us with 190 stars. To determine the membership status of each of the remaining stars, we downloaded proper motion data from the PPMXL catalog \citep{roeser10}. A total of 32,637 stars in the PPMXL catalog are located within one degree of the center of M\,67. In order to use these proper motion values to decide on cluster membership, we cross-referenced the PPMXL stars with the stars from Table 2 in Paper I; we found 258 matches with members and 186 matches with nonmembers. Most (95 per cent) of cluster members occupy a box in proper motion space bracketed by -18 to -2 mas/yr in RA and -12 to -1 mas/yr in DEC. However, 27 per cent of the nonmembers are classified as members with these criteria. Thus, stars classified as nonmembers by this method are very likely to be actual nonmembers, but those classified as members are less certain. Using these proper motion data, we deleted an additional 60 stars, leaving 130 stars.

Next, we removed 7 stars listed as ``BM'' (binary members) in Table 2 of Paper I. As a final constraint on membership status, we also employed the photometric data provided in the PPMXL catalog. The catalog includes $R$ and $K$ magnitudes for all the stars in our working sample. From these, we plotted a $R$ versus $R-K$ color-magnitude diagram and removed giants and subgiants and a few other stars clearly offset from the nominal main sequence. These steps leave us with our final sample of 98 single main sequence M\,67 members.

We calculated the uncertainties in the periods using the following procedure. First, in most cases we set the uncertainty equal to the standard deviation of the mean between the optimal and PDM-derived periods. We adopted a minimum value of 0.5 d. In those cases where the optimal period was multiplied by two (see Section 2.5), the standard deviation is calculated from the adjusted optimal period and the PDM period.

Since many of the stars in our sample fall outside the regions covered by recent photometric studies of M\,67, they lack $V$ and $B-V$ determinations. Cross-matching the stars in our final sample with the ninth data release of the extensive APASS\footnote{http://www.aavso.org/apass/} photometric catalog \citep{hm14}, we are able to obtain $V$ and $B-V$ values for only a subset of them. As a workaround, we calibrated simple equations relating the $R$ and $R-K$ values to the available APASS $V$ and $B-V$ values in order to calculate the later quantities from the former; we used data from 61 stars in the M\,67 field to calibrate the equations. In particular, $V$ is calculated from an equation linear in $R$ and quadratic in $R-K$; $B-V$ is calculated from an equation quadratic in $R-K$. The calibrations have a typical error of 0.11 magnitude, which is about double the typical uncertainty in the APASS $V$ magnitude and the same as that of the $B-V$ magnitude. These calibrations give us $V$ magnitudes and $B-V$ colors for all the stars in our sample. Although other sources of $V$ and $B-V$ photometry are available for many of these stars, we employ only the calculated photometry in this work to maintain homogeneity. The resulting extinction- and reddening-corrected color-magnitude diagram is shown in Figure 1.

\begin{figure}
\includegraphics[width=3.5in]{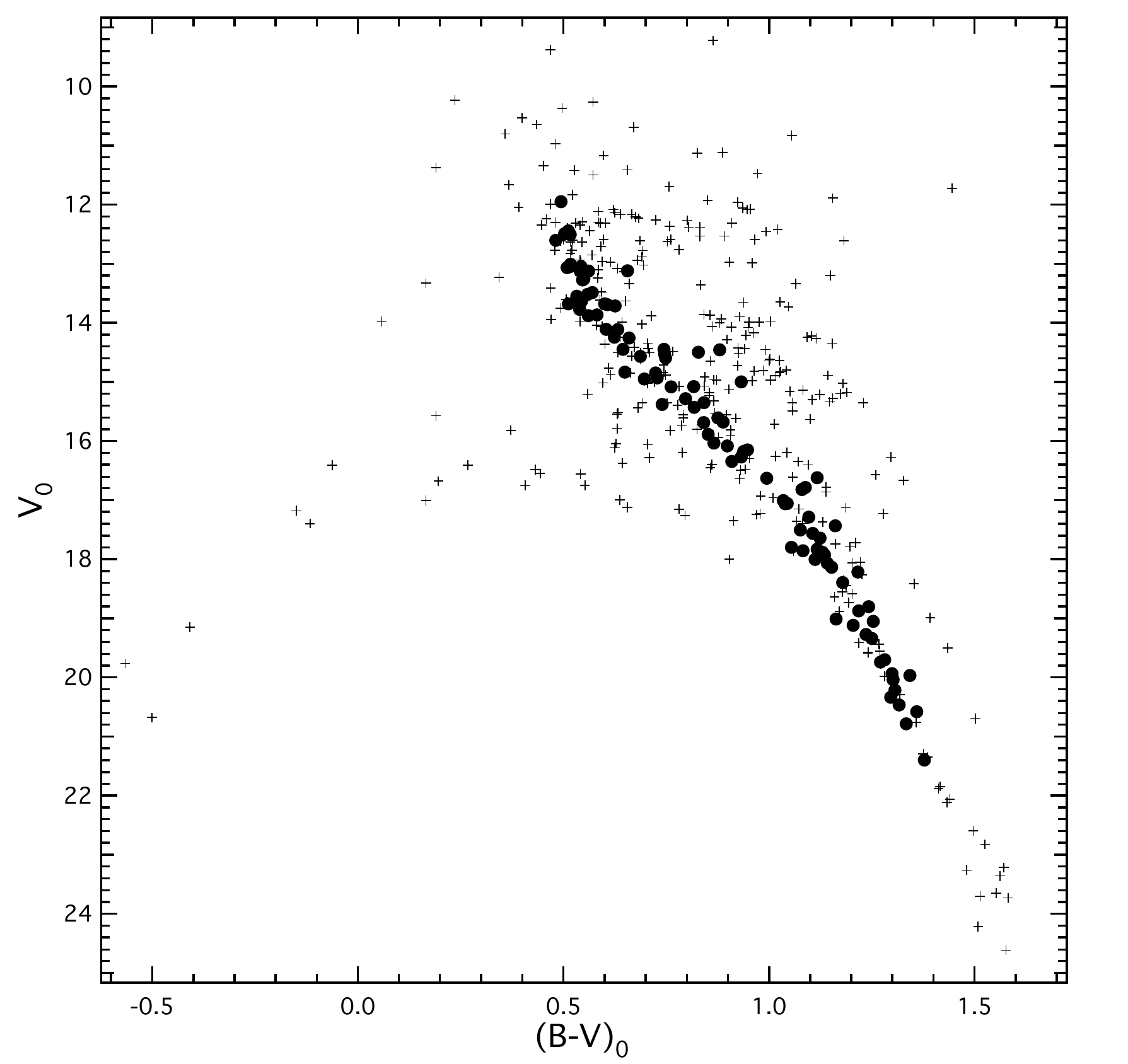}
\caption{Reddening-corrected color-magnitude diagram of the final sample of M\,67 main sequence member stars (dots). The assumed $B-V$ color excess is 0.041 magnitude \citep{tay07}. All the other stars from the original sample in the ``rot'' category are shown as plus signs.}
\end{figure}

We list our period estimates for our final sample in Table 1 and show their distribution on the sky in Figure 2.\footnote{For descriptions of the other calculated quantities listed in Table 1, see Paper I.} We show the period-color diagram in Figure 3.

\begin{figure}
\includegraphics[width=3.5in]{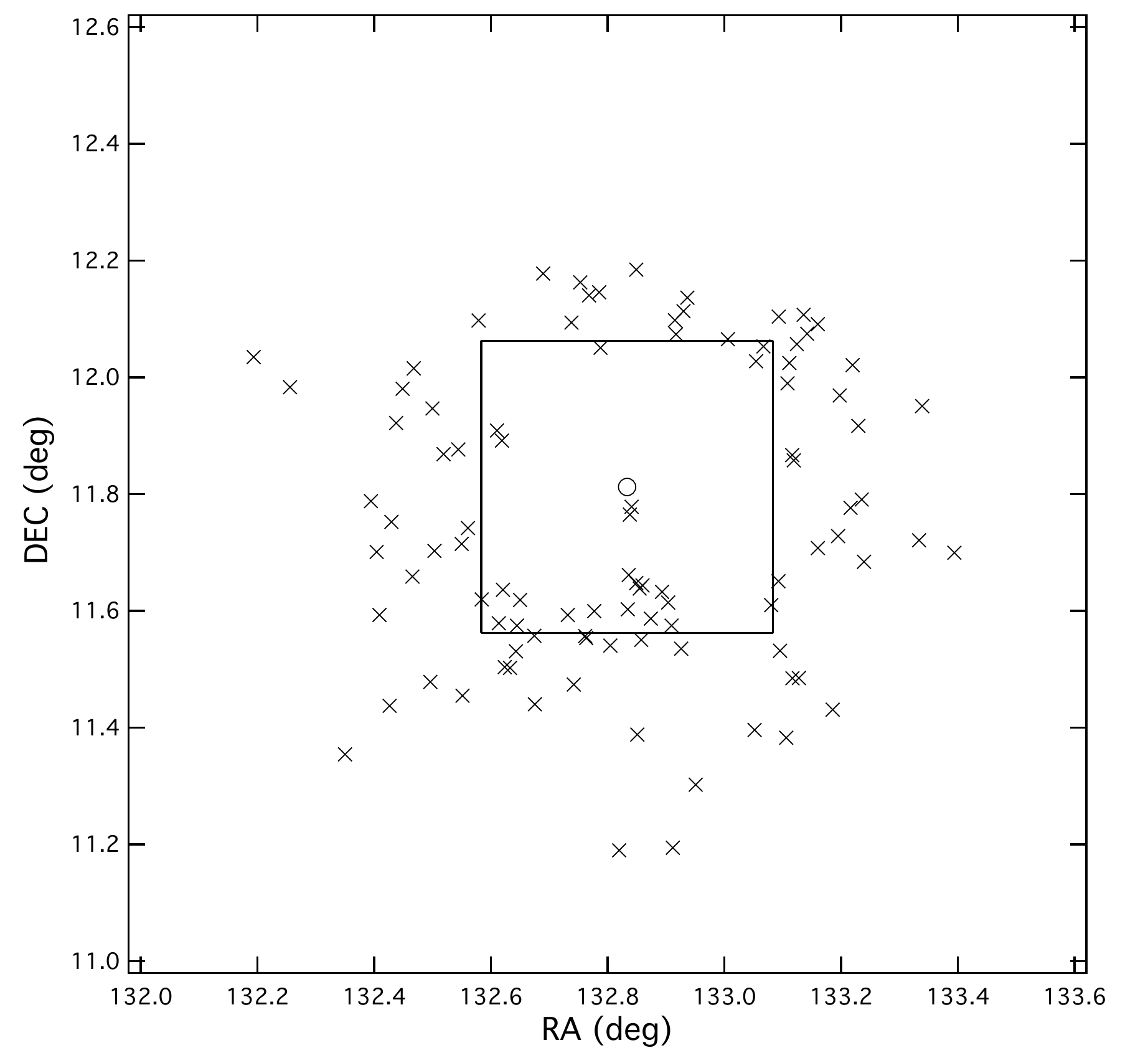}
\caption{Locations of the stars in our final PDCSAP sample on the sky. The center of the cluster is marked with an open circle. The square spans half a degree on a side. North is up and east is to the left.}
\end{figure}

\begin{figure}
\includegraphics[width=3.5in]{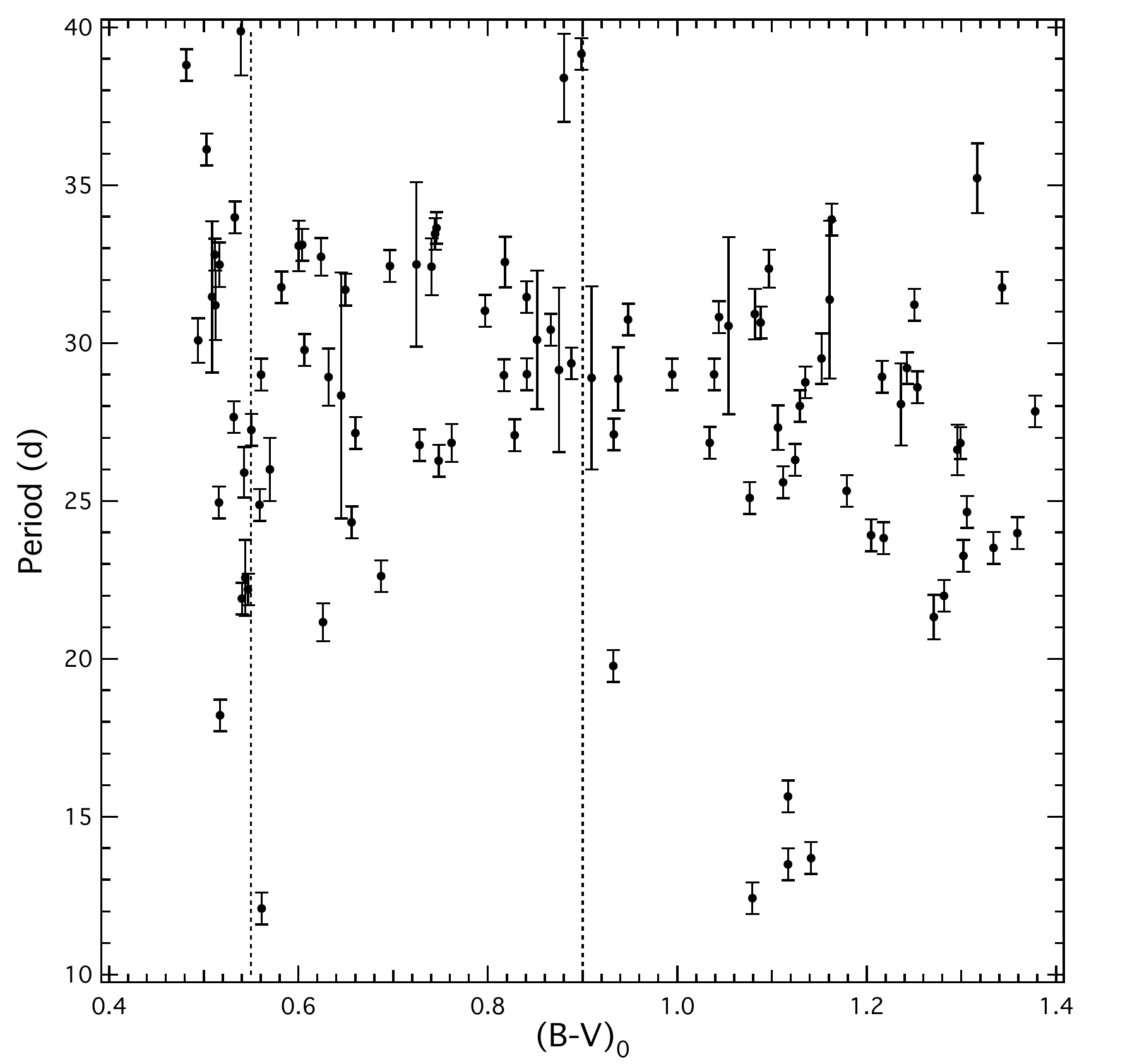}
\caption{Period-color diagram for the stars in our PDCSAP final sample. The magnitudes and colors were corrected for extinction and reddening with an assumed color excess is 0.041 magnitude \citep{tay07}. The two vertical dashed lines bracket the color range $0.55 \le (B-V)_{\rm 0} \le 0.90$. The error bars are also shown on the period values.}
\end{figure}

\begin{table*}
\centering
\begin{minipage}{160mm}
\caption{Derived periods from the PDCSAP\_FLUX values. The complete table is available as on online supplement.}
\label{xmm}
\begin{tabular}{lccccrcccc}
\hline
EPIC & $V$ & $B-V$ & Period & uncer. & FVI & Variability & Corrected variability & RA (2000.0) & DEC\\
 & & & (d) & (d) & & (mmag) & (mmag) & (deg) & (deg) \\
\hline
211371946 & 15.81 & 0.93 & 29.4 & 0.5 & 2.13 & 1.35 & 1.19 & 132.819975 & 11.189986\\
211372217 & 13.65 & 0.60 & 24.9 & 0.5 & 2.20 & 0.39 & 0.34 & 132.912123 & 11.194156\\
211378792 & 14.70 & 0.73 & 22.6 & 0.5 & 2.29 & 0.70 & 0.63 & 132.951537 & 11.302170\\
\hline
\end{tabular}
\end{minipage}
\end{table*}

\subsection{K2SC light curves}

Several 'High Level Science Products' (HLSP) for the {\it K2} mission have been made available recently on the NASA MAST web site.\footnote{https://archive.stsci.edu/hlsp/index.html} These include light curves that have been corrected for systematic photometric variations caused by the Kepler observatory's slow drift along the ecliptic. One of these HLSPs is the collection of detrended light curves produced by \citet{aigrain16}, called K2SC. The raw light curves have been separated into components including detrended fluxes, position-dependent trends and temporal trends. The latter component contains astrophysical variability information. For our analyses, then, we added back the temporal trends to the detrended (PDC-MAP) fluxes following the procedure prescribed by the authors.

Following the same procedure as described in the previous section, we determined the category and period for each light curve. We list the period for each star in Table 2 and show the color-magnitude diagram in Figure 4. We show the period-color diagram in Figure 5. Our final sample contains only 40 stars, less than half as many stars as our PDCSAP sample above; 27 stars are in common between the two sets of results. The reason for this difference is that fewer stars in the K2SC sample fall in the ``rot'' category. This could be caused by, for instance, residual power at certain periods due to imperfect removal of instrumental systematics.

\begin{figure}
\includegraphics[width=3.5in]{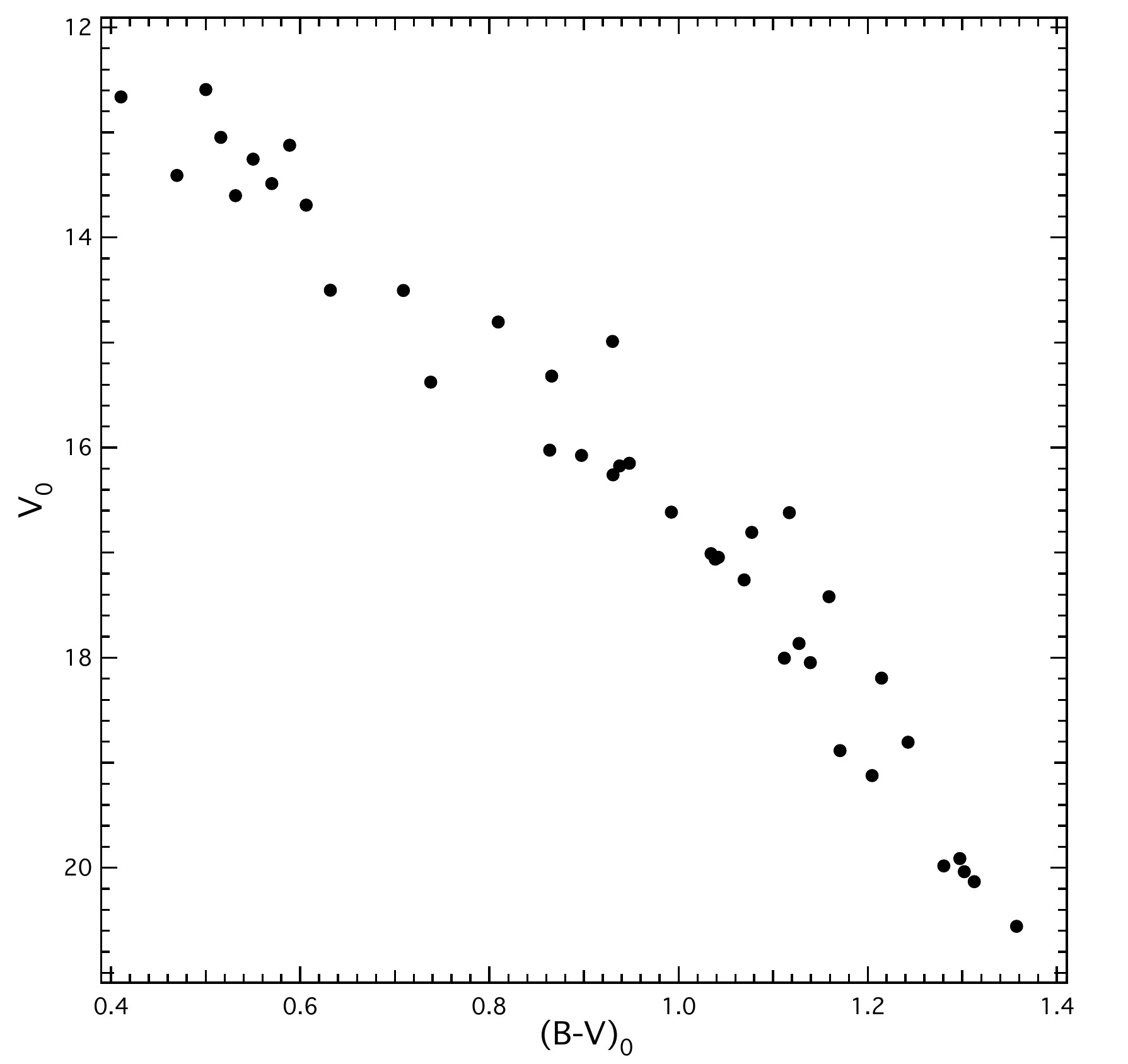}
\caption{Reddening-corrected color-magnitude diagram of the M\,67 main sequence member stars in the K2SC final sample.}
\end{figure}

\begin{figure}
\includegraphics[width=3.5in]{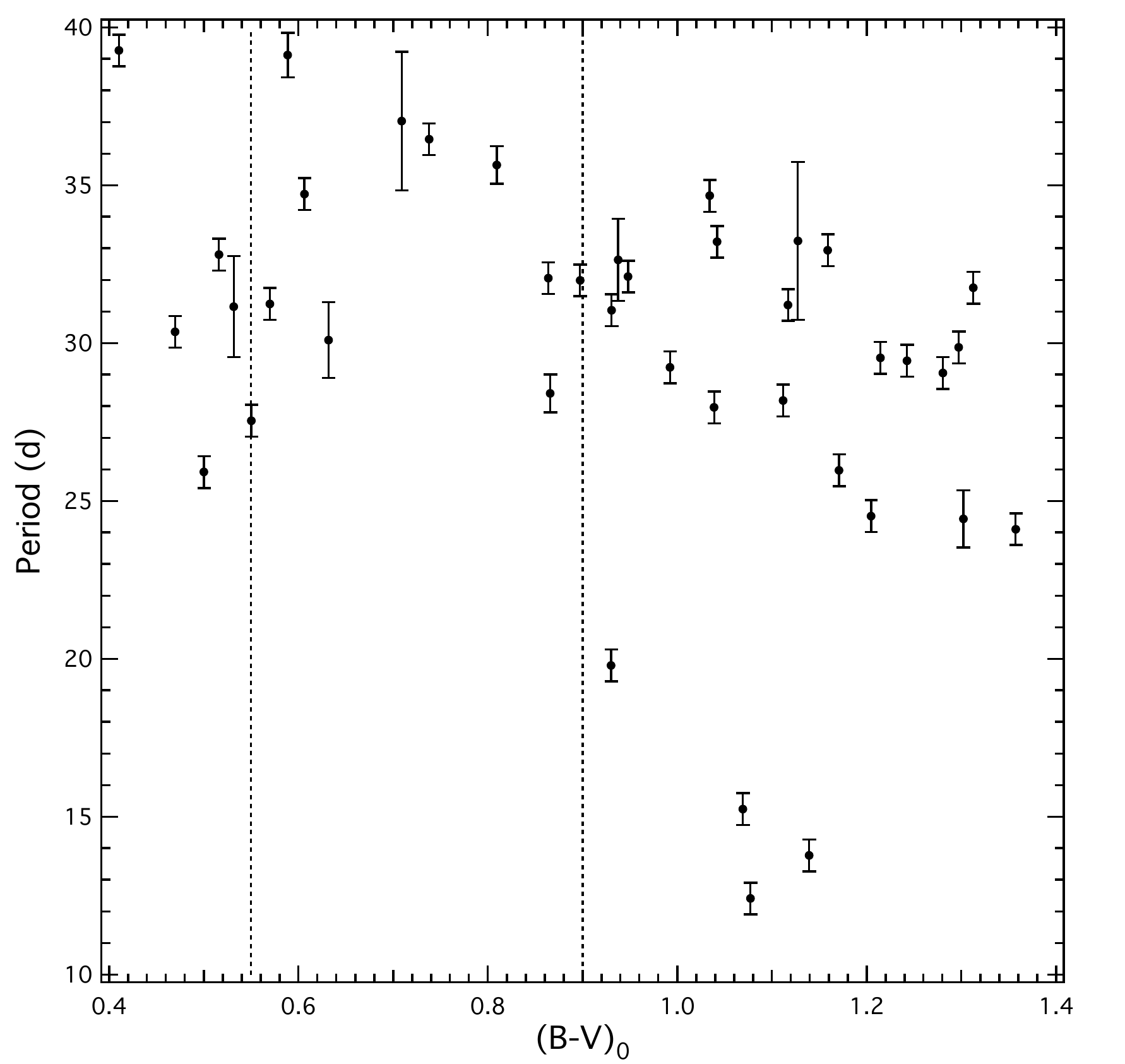}
\caption{Period-color diagram for the stars in our final K2SC sample. All else as in Figure 3.}
\end{figure}

The average difference in periods for the stars in common in the K2SC and PDCSAP final samples is $2.8 \pm 3.5$ d (in the sense of K2SC minus PDCSAP). The star with the largest difference, EPIC 211424980, has a period determined from the PDCSAP light curve almost exactly half the period based on the K2SC light curve. If we exclude it from the statistics, the average difference drops to $2.4 \pm 2.4$ d. We will argue in Section 3.2 that the longer period is the correct one for EPIC 211424980. The three stars with periods shorter than 20 d agree closely. The following linear equation relates the periods for the stars with longer periods:

\begin{eqnarray*}
P_{\rm PDCSAP}=12.0 + 0.526 P_{\rm K2SC}
\end{eqnarray*}

\begin{table*}
\centering
\begin{minipage}{160mm}
\caption{Derived periods from the K2SC light curves. The complete table is available as on online supplement.}
\label{xmm}
\begin{tabular}{lccccrcccc}
\hline
EPIC & $V$ & $B-V$ & Period & uncer. & FVI & Variability & Corrected variability & RA (2000.0) & DEC\\
 & & & (d) & (d) & & (mmag) & (mmag) & (deg) & (deg) \\
\hline
211383622 & 15.12 & 0.97 & 19.8 & 0.5 & 3.10 & 1.16 & 1.10 & 133.106585 & 11.382341\\
211384351 & 14.63 & 0.75 & 37.0 & 2.2 & 2.11 & 0.66 & 0.58 & 133.056695 & 11.393373\\
211386568 & 12.79 & 0.45 & 39.3 & 0.5 & 2.13 & 0.22 & 0.19 & 133.501617 & 11.427949\\
\hline
\end{tabular}
\end{minipage}
\end{table*}

\subsection{Other K2 HLSPs}

Two other HLSP light curve databases are available on the NASA MAST web site: EVEREST \citep{lug16} and K2SFF \citep{vj14}. We downloaded the light curves and attempted to determine periods for the stars in the M67 field from each one, but both are plagued by systematics. In both cases nearly all the stars we examined (a subset of our full sample) displayed nearly identical periodograms, with peaks near 25 and 37 days and weaker peaks at shorter periods. Apparently, the light curves in these databases still contain significant low frequency power that is unrelated to intrinsic stellar variability.

\subsection{Solar trials}

The Sun affords us a unique case of an old main sequence star with accurately known rotation properties. It is instructive, therefore, to test our period analysis methods on solar irradiance data. In particular, we can adjust our methods to minimize the differences between derived and known solar rotation periods.

In Paper I we conducted a large number of automated solar rotation period analyses using VIRGO irradiance data. The VIRGO light curves were prepared in such a way as to mimic the {\it K2} light curves. The resulting mode value of the derived periods was close to the known average synodic solar rotation period, implying that the average rotation period of a group of sun-like stars can be determined reliably from high precision photometry spanning only about 80 days. However, there is room for improvement in our analysis. In particular, we can include subjective input in categorizing each solar periodogram, as we did in Paper I for the M\,67 stars.

Following our procedure described in Paper I, we begin by deriving solar rotation periods from the VIRGO total solar irradiance data from the {\it Solar and Heliospheric Observatory} (SOHO) spacecraft. We downloaded the latest available data,\footnote{ftp://ftp.pmodwrc.ch/pub/data/irradiance/virgo/TSI/virgo\_tsi\_h\_v6\_005\_1602.dat} which includes observations from January 28, 1996 through February 14, 2016. Next, we prepared 400 trial light curves, with the starting time of each one randomly selected from this time span. Additional details of the data preparation are described in Paper I. The optimal period was extracted from each sample light curve in an automated way using gatspy. We also determined the period using the PDM method. Finally, we assigned each periodogram to a subjective category. We had not done these last two steps with the solar trials in Paper I. The most helpful category is the``rot'' category, which corresponds to periodograms with a single dominant peak.

For each solar trial, we also calculated the expected solar synodic rotation period, based on the average heliographic latitude of sunspots at that point in the sunspot cycle. The periods ranged from 26.2 to 29.1 d for the time period covered by the VIRGO data. If the M\,67 sun-like stars exhibit similar rotation properties as the Sun, then we should expect a comparable range in their rotation periods, even assuming insignificant measurement error. This follows because the M\,67 stars will be at various phases in their activity cycles.

The average optimal and PDM periods for all the solar trials are $25.3 \pm 8.4$ and $39.0 \pm 9.5$ d, respectively. This large difference in these mostly results from the presence of PDM periods that are double the optimal periods. There are also a significant number of PDM periods equal to 50 days, which is the maximum period limit in the PDM analysis; these PDM periods are therefore not actual successful period determinations. Excluding the 50 d period values, the average PDM period drops to $35.7 \pm 8.3$ d. In addition, in some of the trials within the ``mult'' category the optimal period was about half the actual synodic period.

The average difference between the optimal and actual solar synodic periods for all the trials is $-2.2 \pm 8.5$ d, while this difference increases to $11.5 \pm 9.4$ d for the PDM periods ($8.2 \pm 8.2$ d excluding the 50 d period values). The corresponding differences drop to $-0.2 \pm 3.2$ and $6.9 \pm 9.9$ d ($2.6 \pm 6.1$ d excluding the 50 d period values) for the ``rot'' category and $-3.1 \pm 9.8$ and $13.6 \pm 8.4$ d ($10.8 \pm 7.8$ d excluding the 50 d period values), respectively, for the other categories combined (mostly ``mult''). 

Of the 400 solar trials we conducted, 123 of them (31 percent) fall into the ``rot'' category. The average optimal and PDM periods of these 123 trials are $27.1 \pm 2.9$ and $34.2 \pm 9.9$ d ($29.9 \pm 6.2$ d excluding the 50 d period values), respectively. In most cases the optimal and PDM periods agree to within 1 day. Most of the difference in the average values between these two sets of periods is due to the presence of a number of PDM period values greater than 40 d. From these results, we can conclude that the optimal periods in the ``rot'' category give the most accurate periods and the smallest scatter. Still, the PDM periods are helpful in some cases, as we explain below.

Given these results, we applied the following rules to each pair of solar trial period values to arrive at a best ``compromise period.'' First, if the PDM period is greater than 40 d, it is deleted; in this case, the best period is set equal to the optimal period. We chose 40 d as the cutoff, because it is about half the duration of the time series (permitting us to satisfy the Nyquist sampling criterion). If the two period values for a given trial differ by less than 5 d, they are averaged. If they are more than 5 d apart and the ratio of optimal period to the PDM is less than 0.7, then the optimal period is multiplied by 2 and averaged with the PDM period; otherwise, they are averaged. This step fixes periods that are half the true value within a range that accounts for scatter. Following this procedure, we were able to significantly reduce the scatter in the periods and also bring the average value of the compromise periods into closer agreement with the actual synodic solar rotation period. Note, however, that we are not claiming that this procedure is the best one to arrive at reliable periods in all cases, but only that it significantly reduces the differences between the derived and actual solar periods (and presumably also for sun-like stars).

For the overall sample, the average difference between the compromise and expected periods is $-0.2 \pm 7.4$ d. For the ``rot'' category, the average difference is $-0.1 \pm 3.2$ d. Thus, even with these corrections, the period estimates for the ``rot'' category remain preferable to those in the other categories. 

We have applied the above procedure to the M\,67 data with two changes. First, if either the optimal or the PDM period is greater than 40 d, then the star is excluded from the final sample. A star is also excluded if one of the periods less than 10 d. The categorization of the periodograms and light curves is a subjective process. In some borderline cases, a light curve might be placed in either the ``rot'' or the ``mult'' category. Therefore, it is helpful to have a method that can extract a reliable period for either category. Still, the results from the solar trials imply that we will only be able to derive reliable rotation periods for a minority of the sun-like stars in M\,67. 

Some of the periodograms of the M\,67 stars will differ from the solar ones due to the presence of other periodic processes not present in the solar photometry. These include the presence of eclipsing binaries and planet transits as well as pulsational variations. Most of these will be obvious from examination of the periodograms and especially the light curves and, therefore, should not enter into the ``rot'' category.

\section{Discussion}
\subsection{Comparison with Previous Works}

Although in the present work we include a larger search area for cluster members around M\,67 compared to Paper I, our final sample from the same PDCSAP light curves includes fewer member stars in the ``rot'' category (98 versus 129). This follows because we have been more restrictive in the present work as to which periodograms we place in the ``rot'' category. For example, the sample periodograms of stars in the ``rot'' category shown in Figures 10 and 11 in Paper I  would have been placed in the ``mult'' category in the present work and therefore not included in our final list in Table 1.

We compare our period estimates from Paper I and the present work to those of B16 in Table 3. All but one of the 20 stars measured by B16 were included in Paper I, but only 9 stars overlap with our new PDCSAP sample and 2 overlap with the K2SC sample. Again, this smaller overlap is due to several stars having been classified in the ``rot'' category in Paper I, which are now categorized as ``mult.'' While there is good agreement between B16 and Paper I for about half the stars, most of the other period estimates differ by a factor of two. The agreement is much better between B16 and the results of the present work; the average difference (us minus B16) is only $-0.2 \pm 1.5$ d for the new PDCSAP sample.

We have been more restrictive in selecting the stars to include in our final sample in Table 1, resulting in a smaller sample size compared to Paper I. This strategy largely avoids the ambiguity resulting from those cases where the true rotation period is not clear cut. This also avoids having to choose between the period estimates of B16 or those of Paper I when they differ by a large factor (often a factor of two). Our latest results lead us to believe that the apparent half-period values from Paper I are probably in error.

Looking at Figure 12 of Paper I, wherein periods are plotted against dereddened colors for the sun-like member stars, we note that the periods at a given color appear to be bimodal. One group averages near 15 days, while the other averages near 25 days. We can see that the shorter period group is much less populous in the new version of the figure (Figure 3). Still, we are not justified in removing these stars from the plot.

\citet{ds16} determined vsini values for 82 nearby single and binary solar twins. Amongst the single twins near the solar age, vsini ranges from about 0.7 to 2.0 km~s$^{\rm -1}$. They also found that binary twins have vsini values up to about 4 km~s$^{\rm -1}$. It is unlikely that this spread in vsini values is due to the range in inclination angles we observe. It implies a real spread in rotation periods at a given age.

There is also independent evidence for differences in the rotation periods among the M\,67 sun-like stars. \citet{RG09} studied the chromospheric activity levels and vsini values of 15 sun-like stars in the cluster. They found all but two of the stars in their sample to have sun-like activity and slow rotation. One of them, S747, is a spectroscopic binary. They report a vsini value of $4.0 \pm 0.5$ km~s$^{\rm -1}$ for the other one, S1452; this is about twice the typical value of the other stars they measured. While they didn't have any evidence for binarity in the case of S1452, \citet{gell15} do list it as a spectroscopic binary. Still, these observations demonstrate that some sun-like stars in M\,67 are fast rotators. It is possible, then, that a few stars in our sample are unrecognized spectroscopic binaries, which could be a reason for some of our anomalously short period estimates.

\begin{table}
\caption{Comparisons of our period estimates with B16}
\label{obs}
\begin{tabular}{@{}lcccc}
\hline
EPIC & P (B16) & P (Paper I) & P (PDCSAP) & P(K2SC)\\
 & (d) & (d) & (d) & (d)\\

\hline
211388204 & 31.8 & -- & 31.7 & -- \\
211394185 & 30.4 & 24.8 & -- & -- \\
211395620 & 30.7 & 30.6 & 30.4 & 32.1\\
211397319 & 25.1 & 24.4 & 28.3 & --\\
211397512 & 34.5 & 16.6 & -- & --\\
211398025 & 28.8 & 14.6 & -- & --\\
211398541 & 30.3 & 27.9 & 30.1 & --\\
211399458 & 30.2 & 16.0 & -- & --\\
211399819 & 28.4 & 26.3 & 26.8 & --\\
211400500 & 26.9 & 25.9 & 26.3 & --\\
211406596 & 26.9 & 26.5 & 26.8 & --\\
211410757 & 18.9 & 19.4 & -- & --\\
211411477 & 31.2 & 14.7 & -- & --\\
211411621 & 30.5 & 15.2 & -- & --\\
211413212 & 24.4 & 24.3 & -- & 30.1\\
211413961 & 31.4 & 37.7 & -- & --\\
211414799 & 18.1 & 9.1 & -- & --\\
211423010 & 24.9 & 44.8 & -- & --\\
211428580 & 26.9 & 25.6 & -- & --\\
211430274 & 31.1 & 26.6 & 29.2 & --\\
\hline
\end{tabular}

\end{table}

\subsection{Gyrochronological Age of M\,67}

B16 derive a gyro-age of $4.2 \pm 0.2$ Gyr for M\,67 from their measurements of the rotation periods of 20 main sequence cluster members. In addition, they derive an age of $4.1 \pm 0.23$ Gyr from published chromospheric activity measurements. Ages based on isochrone-fitting to M\,67 photometry yield ages ranging from about 3.0 to 4.8 Gyr \citep{yad08,sara09}.

Following our procedure in Paper I, we can derive a gyrochronological (gyro-) age for M\,67 from the average period of the sun-like member stars in our final sample. There are 37 stars between the two vertical lines in Figure 3, which demarcates the same range of colors we used in Paper I. For these stars, the mean period is $29.3 \pm 4.8$ d ($\pm 0.8$ d standard error of the mean). This standard deviation value is only moderately larger than that of the solar trials above in the ``rot'' category, implying that most of the scatter in the period determinations is accounted for if we assume these stars behave like the Sun. 

A similar analysis for the 11 sun-like stars in Figure 5 yields a mean period of $33.1 \pm 3.7$ d ($\pm 0.8$ d standard error of the mean). If we adjust the K2SC periods to be on the same scale as the PDCSAP periods with the equation in Section 2.3, we obtain $29.4 \pm 2.0$ d ($\pm 0.6$ d standard error of the mean).

In Paper I we calculated the mean period of 28 single sun-like stars to be $23.4 \pm 6.5$ d ($\pm 1.2$ d standard error of the mean). Clearly, our revised method of period determination has had a very significant effect on the mean value of the period.

B16 report rotation periods along with errors in Table 1 of their study. A simple mean of the periods of the 13 sun-like stars in their sample with $0.55 \le (B-V)_{\rm 0} \le 0.90$ yields a value of $28.3 \pm 2.7$ d ($\pm 0.7$ d standard error of the mean). This is entirely consistent with our mean periods for the same color range.

Given the close agreement between our new PDCSAP periods and the periods reported in B16, it seems likely that our K2SC periods are systematically too high. We can combine these these three datasets if we adjust the K2SC periods using the relation in Section 2.3. Doing so, we have 157 measurements of 121 stars (excluding IDW 4034 from B16, which is classified as an SB1 binary). For EPIC 211424980 we doubled the PDCSAP period, since that would place it in the region where most of the measurements fall and the period would be in agreement with the K2SC value. We show in Figure 6 the individual data values as well and the means and error bars (for data binned in steps of 0.2 magnitude in $(B-V)_{\rm 0}$), and curves for three \citet{barn10} model ages.

\begin{figure}
\includegraphics[width=3.5in]{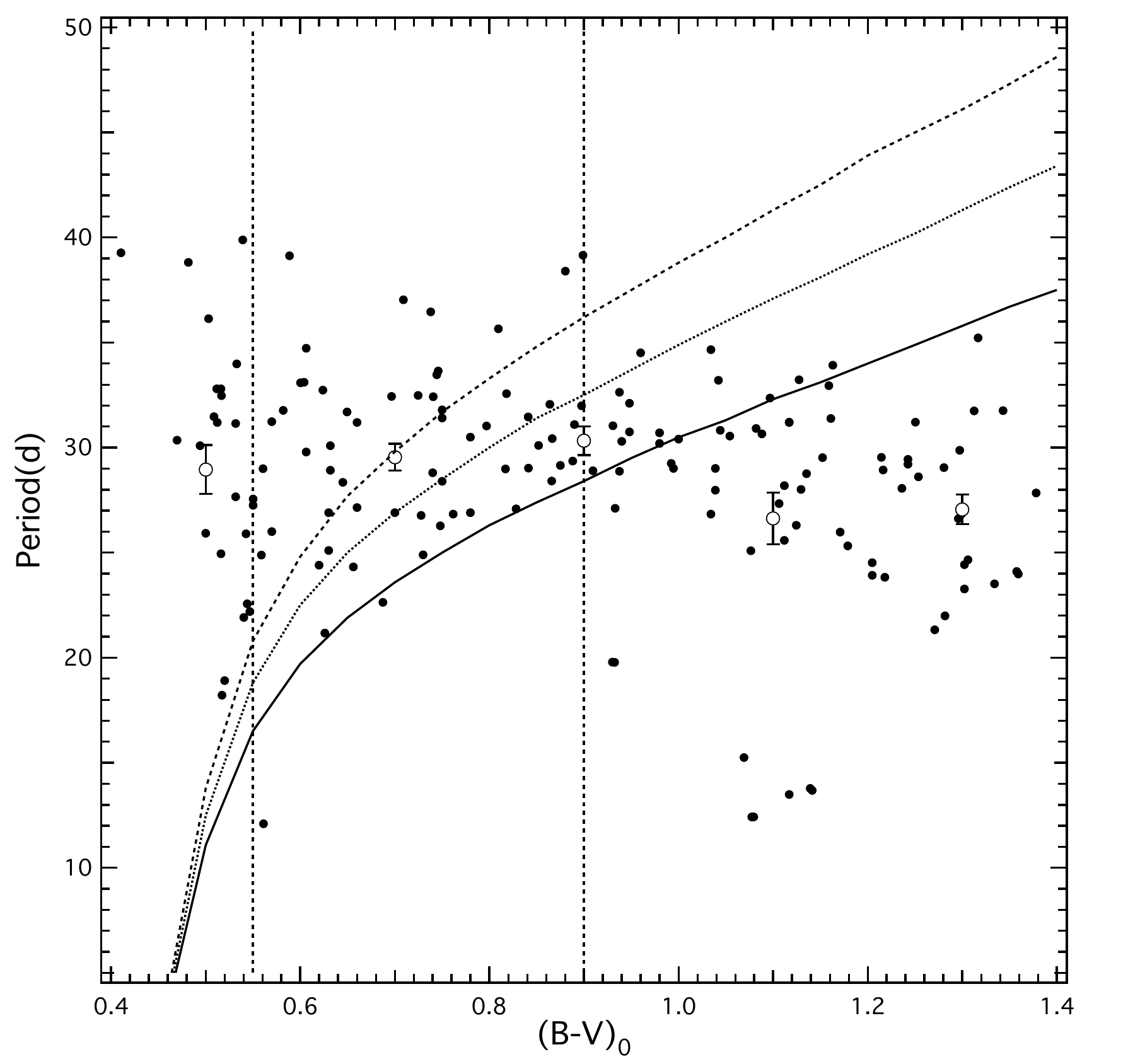}
\caption{Period-color diagram for the combined samples from B16 and the present work. The open circles are the mean values with the error bars indicating standard error of the mean. {\bf The three curves correspond to \citet{barn10} model ages of 3.5 (solid), 4.5 (dotted), and 5.5 (dashed) Gyr.} All else as in Figure 3.}
\end{figure}

The mean period at $(B-V)_{\rm 0} = 0.7$ in Figure 6 is $29.6 \pm 0.6$ d. We can calculate the gyro age of M\,67 based on this period for the sun-like stars. Even before doing the calculation, however, we already know the answer will be greater than solar age, since our average period is larger than the solar value. Following our approach in Paper I, we use the model of \citet{barn10} and \citet{bk10}, which was also used by B16.  We derive a gyro-age of $5.8^{\rm +0.3}_{\rm -0.2}$ Gyr. The gyro-age based on the B16 mean period for the sun-like stars is $5.4 \pm 1.0$ Gyr ($\pm 0.3$ Gyr standard error of the mean).

If, instead, we focus on a more narrow range of $(B-V)_0$, between 0.60 and 0.70 in Figure 6, we calculate a period of $28.7 \pm 3.9$ d ($\pm 0.9$ d standard error of the mean) for 19 stars. This corresponds to a gyro-age of 5.5 Gyr. For the 5 stars in this color range in B16, we calculate an average period of $26.9 \pm 2.6$ d ($\pm 1.2$ d standard error of the mean). For comparison, the solar Carrington sidereal period is 25.4 d.\footnote{It is interesting to note that the gyro-age derived from this sidereal period is 4.3 Gyr. A rotation period of 26.2 d for the Sun would yield an age of 4.6 Gyr.} Again, when we restrict our analysis to the stars most similar to the Sun in M\,67 we find that they rotate more slowly, implying a greater age.

How can we square our new gyro-age estimates with the value B16 derived? There are two possible reasons for this difference. First, the approach of B16 to calculate the average gyro-age of M\,67 is somewhat different from ours. While we choose to focus on the sun-like stars, B16 used all the stars in their sample to arrive at their answer. They calculated individual ages for the stars and then averaged the results. Second, it is not clear from their description of the analysis if they included the uncertainties of the period determinations in their calculations of the average (by weighting each estimate). The star in their sample with the smallest uncertainty in period, IDW 4034, should be excluded from the average. Removing it would raise their quoted age. What's more, two other stars in their sample clearly fall on the binary equal-mass sequence in their Figure 2 (as they note) and should be considered suspect, but they show no evidence of binarity from spectroscopy.

B16 present a color-period diagram (their Figure 4), along with theoretical curves for various ages; their figure is instructive in that it shows the sensitivity of the period-color relation to the age;we show similar curves in Figure 6. From their figure we see that the typical period of main sequence stars in M\,67 should increase from about 23 d at $(B-V)_0 = 0.55$ magnitude to about 32 d at 0.90 magnitude (for their best-fit gyro-age). We do not see evidence of such a trend in Figure 6, either among the individual data points or the binned means. However, assuming the rotation model used by B16 is correct, then some of the spread in our period estimates over this color range is due to real differences in the stars' rotation periods. This would affect the uncertainty of the mean period we calculate for the sun-like stars (perhaps accounting for the larger scatter compared to the solar trials) but not the mean value, since our sample stars are rather evenly distributed across this range of color.

In Table 4 we present the age estimates for the 5 binned mean periods from Figure 6. Clearly, the range in the derived ages is much greater than the quoted error bars, especially for the bluest bin. A simple mean of the 4 redder bins yields an age of 3.5 Gyr. However, since the full dataset does not appear to be consistent with the model of \citet{barn10}, it is not obvious whether we should adopt the mean or a subset. 

We have chosen to adopt the bin with the sun-like stars. First, this bin doesn't have any obvious outliers and displays the smallest scatter in the periods. Second, the sun-like stars are most likely to exhibit sun-like behavior. 

\begin{table}
\caption{Age estimates for the 5 mean binned periods from Figure 6}
\label{obs}
\begin{tabular}{@{}lcc}
\hline
$(B-V)_{\rm 0}$ bin & mean period & Age\\
(mag) & (d) & (Gyr)\\

\hline
0.5 & $29.0 \pm 1.2$ & $23.9 \pm 2.0$\\
0.7 & $29.6 \pm 0.6$ & $5.4 \pm 0.2$\\
0.9 & $30.3 \pm 0.7$ & $4.0_{\rm +0.1}^{\rm -0.2}$\\
1.1 & $26.6 \pm 1.2$ & $2.5 \pm 0.2$\\
1.3 & $27.1 \pm 0.7$ & $2.2 \pm 0.1$\\
\hline
\end{tabular}

\end{table}

In summary, the results of the period analyses of both B16 and the present work imply a gyro-age greater than the solar age by about 1 Gyr for the sun-like stars in M\,67. This is more than 1 Gyr older than the typical age implied by recent stellar evolution models for the cluster \citep{sara09}. It is not clear at this point how to resolve this discrepancy. There are probably sufficient uncertainties about the physics that goes into calculating stellar evolutionary models and the composition of the stars in M\,67 that an age near 5.4 Gyr could be accommodated by them \citep{mag10}.

\subsection{Solar analogs}

In Paper I we newly identified 32 solar analogs in M\,67, which we listed in Table 6. We have identified a few more in the present work. The one most like the Sun is EPIC 211378792; its magnitude and color are very similar to those of S770, which has been identified as a solar twin. Its period is 22.6 d, and its $(B-V)_{\rm 0}$ color is 0.69. In addition, we can add EPIC 211390107 and EPIC 211398269 as solar analogs, with periods of 33.6 and 33.5 d, respectively. These two stars are located very close together on the color-magnitude diagram, which would explain why they have nearly identical periods. The fact that their periods are about 7 days longer than the solar rotation period could be accounted for by their slightly redder colors than the sun, with $(B-V)_{\rm 0} = 0.75$.

\section{Conclusions}

Using {\it Kepler/K2-Campaign-5} light curves, we have performed an improved period analysis of M\,67 member stars compared to our earlier similar study (Paper I). We found that some of the period estimates in Paper I were not true rotation periods. Our new analysis includes stars observed by {\it Kepler} over a wider area than we explored in Paper I, but, because of our more restrictive criteria in the present work, actually yields fewer rotation period determinations from the same data. In addition, we derived periods from a second {\it K2} database containing detrended light curves. The new rotation periods are in excellent agreement with those of B16, where our samples overlap.

We determine the mean period for sun-like stars in M\,67 to be $29.6 \pm 0.6$ d, which implies a gyro-age of $5.4 \pm 0.2$ Gyr. These results are similar to those B16 for their smaller sample of sun-like stars in the cluster. However, we obtain different gyro-ages for the cluster depending on which region of the main sequence we consider.

Progress in the study of rotation periods can be made by increasing the time baseline of {\it K2} observations of M\,67, which would be possible if it is retargeted. Ground-based observations can also be employed for those stars with larger amplitude variations. 

\section*{Acknowledgments}

We thank the reviewer for very helpful suggestions. We acknowledge receipt of the updated dataset (version: 6\_004\_1510) of the VIRGO Experiment, on the cooperative ESA/NASA Mission SoHO from the VIRGO Team through PMOD/WRC, Davos, Switzerland.

\bsp

\label{lastpage}

\end{document}